# Intracellular Electrochemical Nanomeasurements Reveal that Exocytosis of Molecules at Living Neurons is Subquantal and Complex

Anna Larsson[1, +], Soodabeh Majdi[1, +], Alexander Oleinick[2], Irina Svir[2], Johan Dunevall[1], Christian Amatore[2, 3], and Andrew G. Ewing*[1]

¹ Department of Chemistry and Molecular Biology University of Gothenburg, Kemivägen 10, 41296 Gothenburg(Sweden)

² CNRS – École Normale Supérieure, PSL Research University – Sorbonne University, UMR 8640 "PASTEUR", Département de Chimie, 24 rue Lhomond,75005 Paris (France)

³ State Key Laboratory of Physical Chemistry of Solid Surfaces, College of Chemistry and Chemical Engineering, Xiamen University, 361005 Xiamen, China.

[+] These authors contributed equally to this work.

E-mail: andrew.ewing@chem.gu.se

**Abstract:** Since the early work of Bernard Katz, the process of cellular chemical communication via exocytosis, quantal release, has been considered to be all or none. Recent evidence has shown exocytosis to be partial or 'subquantal' at single-cell model systems, but there is a need to understand this at communicating nerve cells. Partial release allows nerve cells to control the signal at the site of release during individual events, where the smaller the fraction released, the greater the range of regulation. Here we show that the fraction of the vesicular octopamine content released from a living *Drosophila* larval neuromuscular neuron is very small. The percentage of released molecules was found to be only 4.5% for simple events and 10.7% for complex (i.e., oscillating or flickering) events. This large content, combined with partial release controlled by fluctuations of the fusion pore, offers presynaptic plasticity that can be widely regulated.

Two works published in 2010 suggested that the Katz principle,[1] was incorrect for all-or-none release and that only part of the chemical load of vesicles was released during exocytosis, at least as measured as a full spike during amperometry.[2] The combination of electrochemical methods to measure both release and vesicle content in 2015 added a wealth of information to support the concept of partial release in exocytosis.[3] Additionally, this has recently been supported by work with TIRF microscopy showing 'subquantal' release from vesicles in adrenal chromaffin cells and using super-resolution STED microscopy.[4] It appears that the full event generally involves release of only part of the load of chemical messenger in single-cell model systems like adrenal chromaffin and PC12 cells. Is this also true at living neurons in a nervous system and to what extent?

To answer this critical question, we quantified the number of octopamine molecules in the neuromuscular neurons of *Drosophila* larvae by adapting an amperometric technique developed in our group to herein determine vesicular content in single living and communicating neurons. This technique, intracellular vesicle impact electrochemical cytometry (IVIEC),[3a] uses a nanometer sharp, needle-shaped, carbon fiber electrode to pierce the plasma membrane of a single secretory cell in order to access the vesicles in the cytoplasm. Here, we pierced an even smaller varicosity along the neuronal axon. Upon application of a potential at the electrode, vesicles that have adsorbed to the surface of the electrode open to the electrode surface via electroporation and the electroactive transmitters inside are quantified using the relationship between the measured charge and mole amount that is described by Faraday's law.[5]

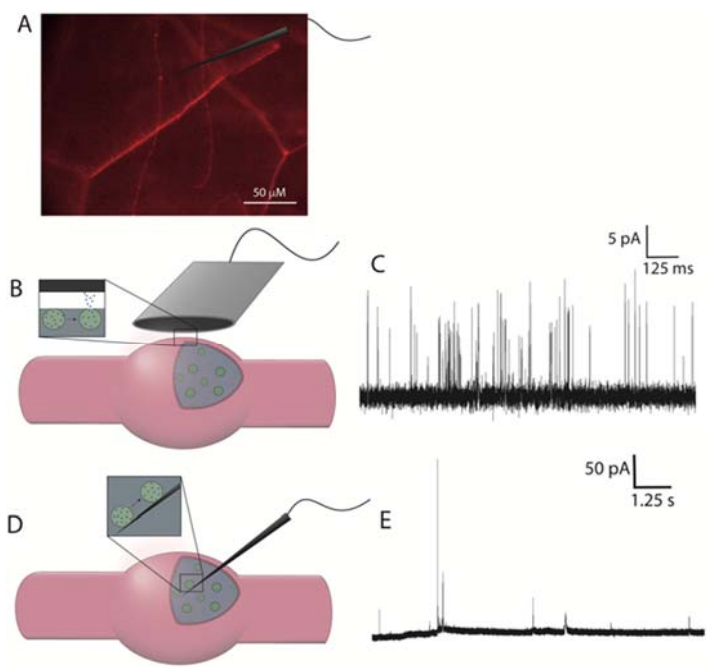

**Figure 1.** A, Fluorescence microscopy image of mCherry labeled neurons in Drosophila larva showing the varicosities and a schematic of a nanotip electrode placed on or in a varicosity. B, Schematic of an electrode on a varicosity with representative exocytosis current transients in C. D, Schematic of a nanotip electrode placed in the varicosity for IVIEC and representative current transients for vesicle content measurements in E.

*Drosophila* third instar larva that express the fluorescent protein mCherry specifically in type II varicosities (octopaminergic terminals; fluorescent images and an electrode schematic shown in Fig. 1A) were chosen for both single cell amperometry (SCA), as shown in Figure 1B with a typical current trace in 1C, and IVIEC, shown in Figure 1D with a representative current trace shown in 1E. To carry this out, we resorted to fluorescence as a guide and a sharp nanotip electrode was used to pierce the muscle into the varicosity, and a potential of 900 mV *vs* Ag|AgCl was applied. Transmitter vesicles inside the cell rupture on the electrode leading to current transients from oxidation of the octopamine content (Fig. 1E). Surprisingly, despite the smaller vesicle size in the neurons, the number of molecules detected by IVIEC (Fig. 1E) was dramatically larger than would be expected from previous work in cell cultures,[3] with a median of 441 000 molecules/vesicle (Table 1) and a range of values generally from 200 000 to 1.3 million molecules/vesicle.

When measuring the amount of released octopamine from the larval neuromuscular junction (NMJ) by SCA (Fig. 1B),[6] we observed a range of different types of events. Approximately half of the exocytotic events were complex in nature, suggesting a flickering fusion pore during release. Most of the rest of the spikes were simple events like those typically observed at model cells and we quantified the number of molecules released for these events at 20 000, a number considerably smaller than the transmitter load found with IVIEC above. To further understand the complex events in terms of the fusion pore behavior that regulates exocytosis,[7,8] mathematical reconstruction was used to gain knowledge about the dimensions of the pore as well as to provide another estimate of the total octopamine content inside the vesicle and further assess the extent to which exocytotic release is partial (or 'subquantal') in this neuronal system.

Amperometry of exocytosis from the larval NMJ was mathematically modelled[8] with an approach that has been confirmed by comparison to extended numerical computations and simulations.[9] The complete reconstruction procedure as well as the assumptions made can be found in the SI. Briefly, the ratio of total electroactive content in the vesicle and the released content was tuned so the pseudo-rate constant of release $k°_{diff}(t)$ (proportional to pore size), fluctuates around a constant value except during the sharp initial rise and final decay of the complex event where $k°_{diff}(t)$ increased or decreased rapidly featuring respectively the fast opening and final closure of the pore. This was done for individual events allowing the total initial amount of octopamine in the vesicle to be determined. By also assuming an average vesicle radius[10] of 45 nm and a diffusion coefficient identical to that measured in rat NMJs,[8a] values of the fusion pore radius could be obtained (0.85-2.9 nm). This procedure was repeated for a total of 110 complex events. For simple events, where the current rapidly increases followed by an exponential decrease, a different approach was used. In order to determine the total vesicular content for these events, fusion pore radii were assumed based on literature values and consistent with the data from complex events. As a result, a range of values for the total vesicular amount could also be theoretically determined for these types of events based on reconstruction of 67 simple events.

Selected release events have been overlaid with the modelled fusion pore dynamics in Figure 2A-D. The flickering behavior of the fusion pore as seen in the fine fluctuations is consistent with previous data obtained from neuroendocrine and hippocampal cells as well as in reconstituted *in vitro* model systems [8a,11]. The median value of the average pore size is 1.3 nm, which agrees well with the value reported by patch-clamp (1.2 nm) for the initial fusion pore radius in chromaffin cells.[12] The median octopamine content per vesicle giving rise to simple or complex events (~240 000 molecules, Table 1) was theoretically predicted based on the above modelling of SCA currents (See also SI). SCA measures the amount released and prior evidence has shown that the vesicle is made up of fast release and more tightly bound pools of neurotransmitter.[13b-d] Although the ranges of the modelled and IVIEC values overlap, the difference in the medians provides interesting comparison. The ratio of the theoretical vesicle content based on SCA to the content measured by IVIEC (441 000 molecules) is 55% suggesting that the model of the release transients is predicting the mole amount of octopamine in the easy to release pool in the vesicle. Interestingly, this ratio is similar to that (50 to 60%)[13a] exhibited by PC12 endocrine cells, again revealing that releasing vesicles store neurotransmitters in two different internal pools release.[13b-d] This interpretation is fully compatible with a single exponential decay of the SCA spikes observed for single events.[13b] Furthermore, this hypothesis is supported by the spatial distribution of catecholamines within PC12 vesicles monitored by TEM and NanoSIMS.[13e]

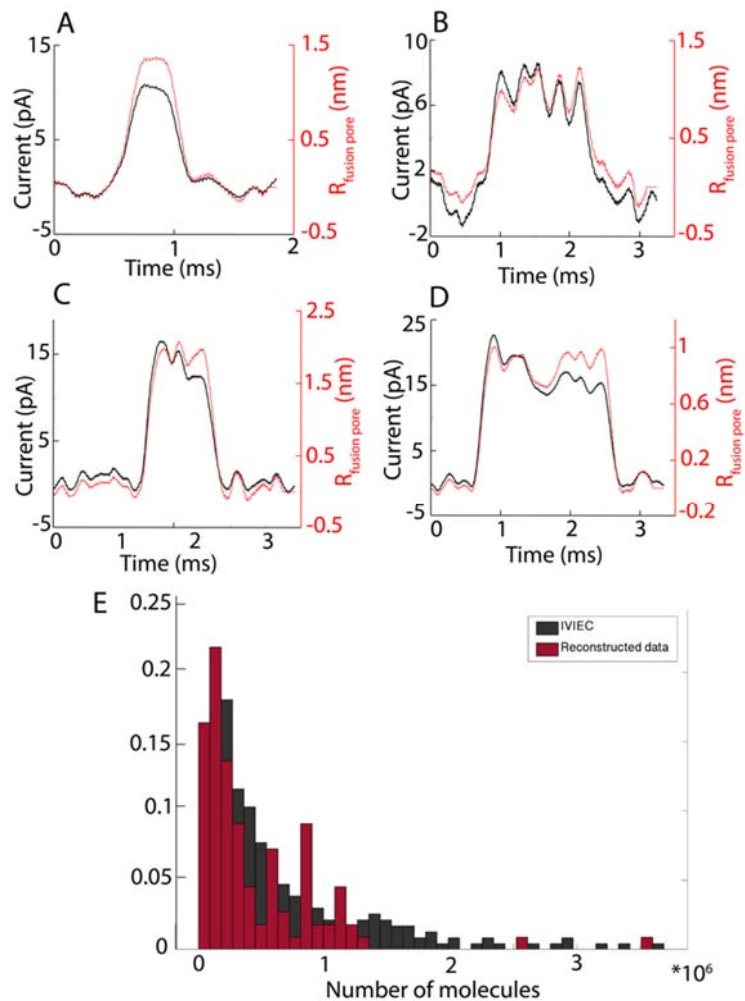

**Figure 2.** A-D, Overlays of exocytosis events measured by SCA (black) and mathematically modelled variations of fusion pore radius (red) for comparison and prediction of intravesicular transmitter content for 4 different events showing simple open and closed exocytosis (A), and different levels of pore opening, flickering and closing during complex (i.e., oscillating or flickering) events (B-D). E, Histograms showing the frequency of vesicle transmitter content measured by IVIEC (black) versus that obtained by mathematical reconstruction (red).

**Table 1.** The number of released versus stored molecules from larva neuromuscular junction vesicles.

| | Simple events ($*10^3$) | Complex events ($*10^3$) |
|---|---|---|
| Molecules released in exocytosis (SCA)[a] | 20 (11-27) | 47 (33-76) |
| Vesicle content reconstructed by modelling SCA events[b] | 237 (175-306) | 240 (121-630) |
| Vesicle content measured electrochemically by IVIEC[c] | 441 (225-919) | |
| Measured release percentage by comparison of exocytosis and IVIEC measurement | 4.5% | 10.7% |

[a] Values given as medians (first quartile-third quartile). [b] Values given as medians (first quartile-third quartile) and were obtained as described in SI. [c] Value given as median (first quartile-third quartile) as was measured at varicosities from n=7 larvae with 232 IVIEC current transients. IVIEC measured a total vesicle content without indication about the nature (simple or complex) of the event that would have prevailed if fusion would have occurred, so we leave this IVIEC value in the middle of the two release modes.

Histograms of the distributions of the number of molecules contained inside NMJ octopaminergic vesicles calculated from the simulations are compared to those measured electrochemically and are shown in Figure 2E. If the average vesicle has 45 nm radius,[10] then for the 441 000 molecules measured, the median intravesicular concentration of octopamine is 1.9 M, illustrating that a mechanism, such as protein binding, must be present to maintain such a high concentration. This concept of protein-transmitter interaction for enhanced vesicular loading has been observed in other secretory cell models such as the adrenal chromaffin cells.[14] By comparing the released amount to the full vesicle content, we can calculate the extent of partial release in this neuronal system. Octopamine release at 20 000 molecules for simple events suggests that during these exocytosis events in the larva NMJ only 4.5% of the transmitter load is released. This number is more around 10.7% for the complex events where 47 000 molecules are released through a series of pore opening/pore closure phases (Table 1).

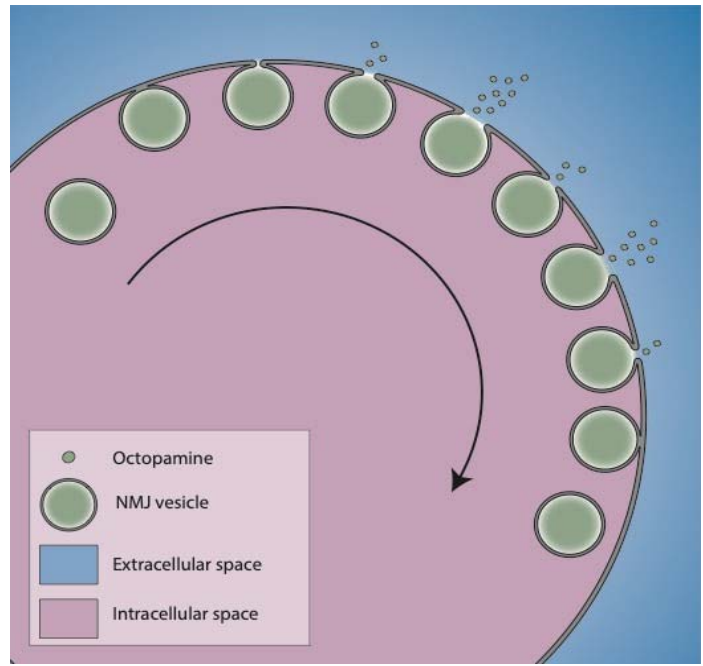

**Figure 3.** Model depicting a complex release event at a neuronal varicosity. The vesicles open and close at the cell membrane once for simple events and flicker for complex events, in both cases only releasing a small fraction of their transmitter load.

The combined results suggest that only a small fraction of the vesicular content is released in the type II neurons of the larva. The number of molecules and percentages for different types of events and methods are detailed in Table 1. For the smallest percentage, <5%, for the so-called simple events, the fusion pore is believed to open and close only once. In comparison, for events where the fusion pore is thought to open, flicker and close (Fig. 3, and Fig 2B-D for pore radius fluctuations), the percentage released increases to nearly 11%, but even here the largest percentage released is considerably lower than that observed in cell culture models (PC12 and adrenal cells) where the percentage released is typically around 60%.[3a, 15]

The importance of partial release of chemical transmitters is that synaptic plasticity can be, at least in part, related to presynaptic regulation of individual exocytotic release events as well as the efficiency of transmitter storage in neuronal vesicles and postsynaptic changes. The fraction of partial release has recently been shown to be plastic in model cells with drugs,[16] and a repeated stimulation paradigm.[17] The highly partial nature of exocytosis in this neuronal system suggests the possibility that cells fine-tune chemical signaling by adjusting the duration and size of the fusion pore opening. The smaller the fraction released the more sensitive the system plasticity will be to a change in release. By increasing the released percentage from an individual simple event from 5% to 10%, the effective transmitter release is doubled. A change of this magnitude could also then only require half of the frequency of events to achieve the same signal across the synapse, allowing an interplay between exocytosis frequency and fraction of transmitter released to regulate plasticity. Thus, partial release and complex pore opening/closing lead us to new theoretical modalities for understanding the initial plastic changes in learning, and also provides a valuable target in drug development for diseases related to early plasticity.

**Acknowledgements**

We acknowledge funding from the European Research Council (Advanced Grant), the Knut and Alice Wallenberg Foundation, the Swedish Research Council (VR), the National Institutes of Health, and CNRS, Ecole Normale Superieure, PSL Research University, and Sorbonne University (UMR 8640 „PASTEUR"). C. A. acknowledges Xiamen University for his position of Distinguished Visiting Professor.